\documentclass[pre,preprint]{revtex4-1}
\usepackage{graphicx}
\usepackage{amsmath,amssymb}
\def\vp{\varphi}
\def\t{\vartheta}
\begin{document}
\title{Chimeras on a social-type network}
\author{Arkady Pikovsky}
\affiliation{Department of Physics and Astronomy, University of Potsdam, 14476 Potsdam-Golm, Germany}
\affiliation{Department of Control Theory, Lobachevsky University of
Nizhny Novgorod, Gagarin Avenue 23, 603950 Nizhny Novgorod, Russia}

\date{...}
\begin{abstract}
We consider a social-type network of coupled phase oscillators. Such a network
consists of an active core of mutually interacting elements, and of a
flock of passive units, which follow the driving from the active elements, 
but otherwise
are not interacting. We consider a ring geometry with a long-range coupling, where
active oscillators form a fluctuating chimera pattern. We show that 
the passive elements are 
strongly correlated. This is explained by negative transversal Lyapunov 
exponents.
\end{abstract}
%
%
\maketitle
\section{Introduction}
Since their discovery about 20 years ago by Kuramoto and 
Battogtokh~\cite{Kuramoto-Battogtokh-02}, chimera patterns
attracted large interest in studies of complex systems.
Chimera is an example of a symmetry breaking in a 
homogeneous system of coupled
oscillators: together with a homogeneous fully synchronous state
there exist non-homogeneous states where some oscillators 
are synchronized and some not. In spatially extended systems
chimera appears as a localized pattern of 
asynchrony~\cite{Abrams-Strogatz-04,Laing-09,Bordyugov-Pikovsky-Rosenblum-10,%
Omelchenko-13,Xie_etal-14,Maistrenko_etal-14,Maistrenko_etal-15,Panaggio-Abrams-15,Smirnov-Osipov-Pikovsky-17,%
Bolotov-Smirnov-Osipov-Pikovsky-18,Omelchenko-Knobloch-19}. In globally coupled
populations chimeras are also possible: they emerge 
not as spatial patterns, rather
a group of asynchronous oscillators ``detaches'' from the synchronous 
cluster~\cite{Yeldesbay-Pikovsky-Rosenblum-14,Schmidt_etal-14,
Schmidt_Krischer-15,Zaks_Pikovsky_17,%
Goldschmidt-Pikovsky-Politi-19}. 

\begin{figure}
\centering
\includegraphics[width=0.3\columnwidth]{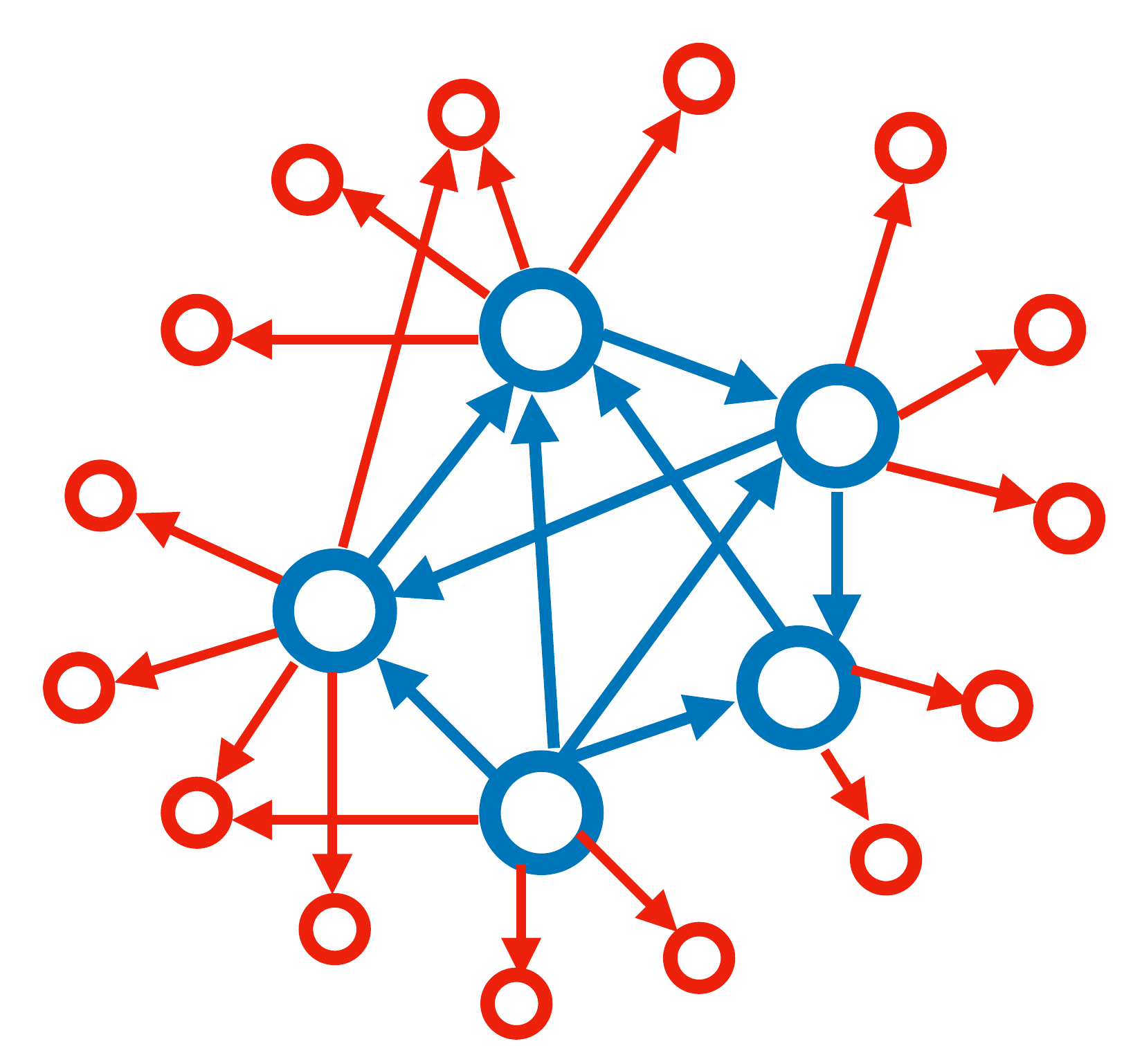}
\caption{Illustration of a social-type network. Central blue
units are active, they interact with each other, but do not get inputs
from passive units (peripherial red ones). The passive units are driven
by the active ones, and do not interact with each other.}
\label{fig:stn}
\end{figure}

The basic model of Kuramoto and Battogtokh is a one-dimensional
ring of phase oscillators with non-local coupling. Each oscillator
is coupled to all others in a symmetric bidirectional way; the strength
of coupling depends on the distance on the ring. There are two 
typical setups for this distance dependence: exponential 
as in~\cite{Kuramoto-Battogtokh-02} (or its modification taking into account
spatial periodicity~\cite{Smirnov-Osipov-Pikovsky-17}), or $cos$-shape 
coupling~\cite{Abrams-Strogatz-04}. In both cases, chimera lives on a symmetric
weighted bidirectional network. This paper aims to generalize the basic
setting of the Kuramoto and Battogtokh to a \textit{social-type network} (STN). 
Such a network, introduced in \cite{Peter-Gong-Pikovsky-19}, deserves a detailed
description. It is a weighted directional network with two types of nodes:
(i) active nodes that force other nodes and potentially are also forced by them 
(i.e., active nodes have outgoing links); (ii) passive nodes that are driven
by active nodes but do not influence them (i.e., passive nodes have only in-going
links). We illustrate this in Fig.~\ref{fig:stn}. The name ``social-type'' is picked
because separation into active and passive nodes is similar to the separation
of social networks into ``influencers'' and ``followers''. The latter participants
get input from the former ones, but not vice versa. In physics,
there are several prominent models of such type. In a restricted many-body
problem in celestial mechanics, one considers several heavy bodies that
interact and move according to the gravitational forces they produce. Additionally, light bodies move in the gravitational field created by the heavy ones but
do not produce gravitational forces themselves (in fact, these forces are neglected
in this setup). Another situation is modeling of two-dimensional turbulence 
by a motion of point vortices~\cite{Eyink-Sreenivasan-06}. The vortices move as 
interacting fluid particles, while other particles, like passive tracers, follow
the velocity field created by vortices but do not contribute to it.

Below we construct the STN by taking a symmetric Kuramoto-Battogtokh network, and
equipping it with additional passive oscillators. We will mainly consider a situation
where the number of passive units is much larger that the number of active ones.
The model will be introduced in Section~\ref{sec:bm}. In Section~\ref{sec:vis}
we will illustrate the dynamics of passive units, and in Section~\ref{sec:st} will
perform its statistical evaluation. 

\section{Basic Model}
\label{sec:bm}
We consider a network consisting of $N$ active phase oscillators $\vp_n$
and $M$ passive phase oscillators $\t_m$. Both are uniformly distributed in space on a
ring $[0,1)$, the coordinates of active units are $x_n=(n-1)/N$, $n=1,\ldots,N$;
the coordinates of passive units are $y_m=(m-1)/M$, $m=1,\ldots,M$. All oscillators
have identical frequency (which we set to zero chosing the appropriate 
rotating reference frame), and are nonlocally coupled:
\begin{align}
\dot\vp_n=&\frac{1}{N}\sum_{k=1}^N G(x_k-x_n)\sin(\vp_k-\vp_n-\alpha),\quad n=1,\ldots,N\;, \label{eq:aosc}\\
\dot\t_m=&\frac{1}{N}\sum_{k=1}^N G(x_k-y_m)\sin(\vp_k-\t_m-\alpha),\quad m=1,\ldots,M\;.\label{eq:posc}
\end{align}
One can see that this coupling implements an STN: while active oscillators are mutually coupled,
passive ones just follow the force from the active ones.

In previous literature, several coupling kernels $G(\cdot)$ has been explored. Kuramoto and 
Battogtokh~\cite{Kuramoto-Battogtokh-02} used 
an exponential kernel, Abrams and Strogatz~\cite{Abrams-Strogatz-04} used a cos-shaped one.
We will follow the latter option, and set $G$ as
\begin{equation}
G(x)=1+A\cos(2\pi x)\;.
\label{eq:kernel}
\end{equation}
Parameters $A=3/4$ and $\alpha=\pi/2-0.05$ are fixed throughout the paper.

Nontrivial properties in the social-type network \eqref{eq:aosc},\eqref{eq:posc}
can be expected if the number of active oscillators $N$ is not too large. Indeed,
in the thermodynamic limit $N\to\infty$ the field created by active oscillators
is stationary (in a certain rotating reference frame), and the dynamics of passive
oscillators in this field is trivial. In contradistinction, for relatively small
$N$ there are significant finite-size fluctuations, which, as we will see, 
lead to nontrivial effects. On the other hand, it is known that chimera in a very
small population is a transient state~\cite{Wolfrum-Omelchenko-11}. Below in this paper we
choose $N=32$; for the parameters chosen 
chimera in Eqs.~\eqref{eq:aosc} is strongly fluctuating 
and has a long life time.

\section{Visualization of Chimera}
\label{sec:vis}

\begin{figure}
\includegraphics[width=0.6\columnwidth]{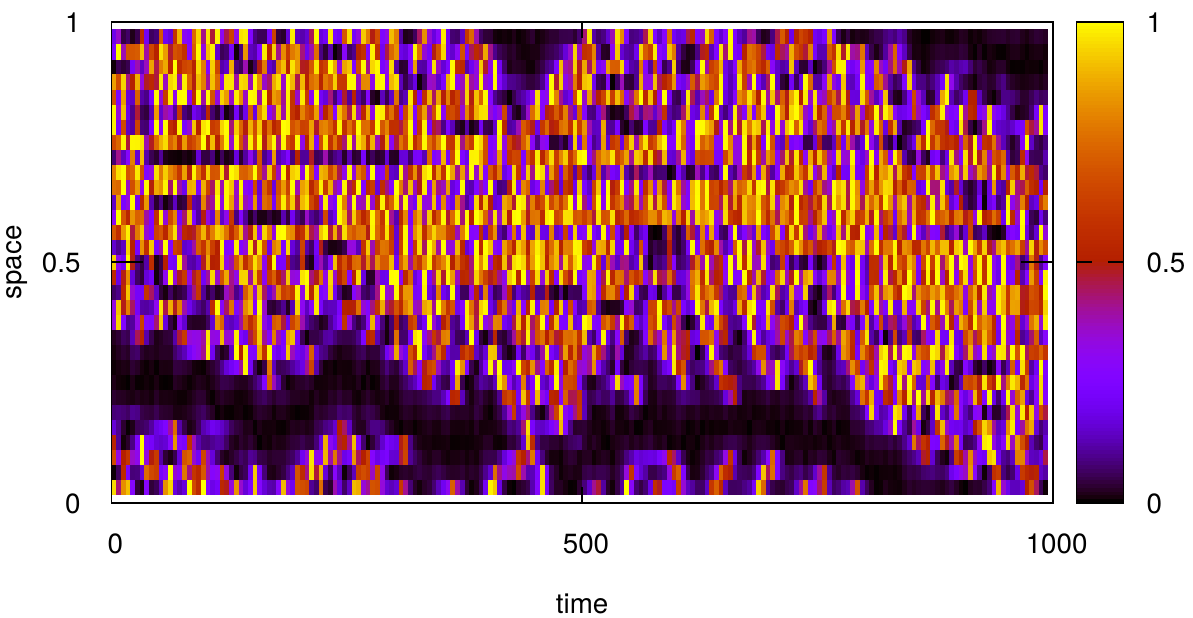}
\caption{Chimera in a set of 32 active units~\eqref{eq:aosc}. Color coding
shows the distances between neighboring units $D_k$, as function of time. Black region
corresponds to a synchronized domain, yellow-red irregular pattern to the desynchronized one.
The position of the synchronized domain experiences  a random walk, so that
the dynamics on the long time scale is ergodic - each oscillators participates 
in synchronous and asynchronous motions.}
\label{fig:vc}
\end{figure}

In Fig.~\ref{fig:vc} we illustrate the chimera state in the set
of active units $\vp_k$. We show the distance between the states
of neighboring active oscillators $D_k=|\sin(\frac{\vp_{k+1}-\vp_k}{2})|$.
This quantity is close to zero if the phases $\vp_k$ and $\vp_{k+1}$
are nearly equal, and is $1$ if the phase difference is $\pi$. In  Fig.~\ref{fig:vc} 
the black region corresponds to a coherent domain of chimera (all the phases
here are nearly equal), and the rest with red/yellow colors is the disordered 
state.

\begin{figure}
\includegraphics[width=0.6\columnwidth]{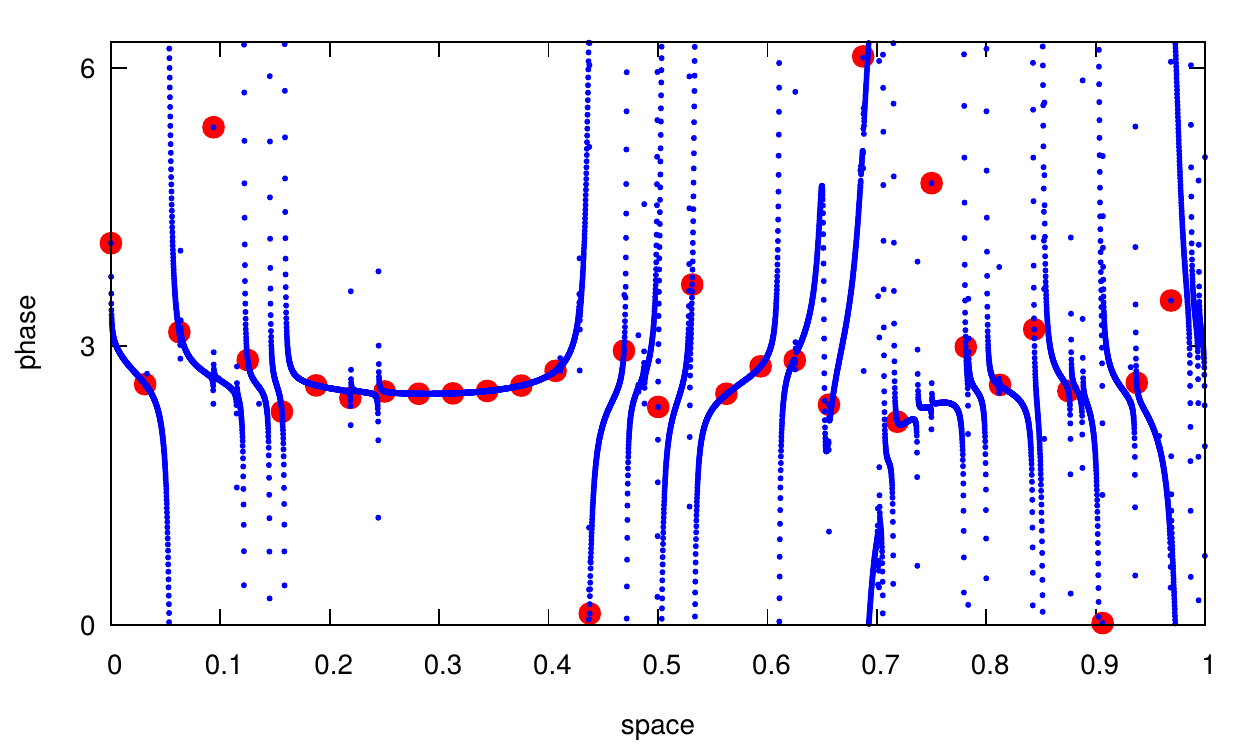}
\caption{A snapshot of an STN with $N=32$ active units (large red filled circles)
and $M=8192$ passive units (blue dots).}
\label{fig:ss}
\end{figure}

Next we illustrate what happens to passive oscillators in the 
regime depicted in  Fig.~\ref{fig:vc}. In Fig.~\ref{fig:ss} we
show a snapshot of the states of active and passive oscillators.
It has following features:
\begin{enumerate}
\item First we mention that the passive elements which have exactly the same
positions as the active ones, attain the same state. This is due to the fact
that although initial conditions are different, these pairs are driven by
exactly the same field, and the conditional Lyapunov exponents are negative (see a detailed
discussion of Lyapunov exponents below), so
that active and passive oscillators synchronize.
\item The active oscillators show typical for chimera domains where the phases are
nearly equal (here $0.18\lesssim x\lesssim 0.4$), and another one,
where neighboring elements do not have close
phases. In contradistinction, close in space passive oscillators 
typically have also close values of the
phases. Visually this appears as a continuous profile of passive phases values. 
Of course, this profile
cannot be exactly continuous because of phase slips, which are also clearly visible
in Fig.~\ref{fig:ss} (e.g., at $x\approx 0.21$ and
at $x\approx 0.24$). Such a phase slip disappears due to finite spacing between passive
elements and the stabilizing role of the negative Lyapunov exponent. Thus, passive
oscillators possess certain degree of regularity also in the domain where active 
oscillators are disordered.
\end{enumerate}

\section{Statistical Properties}
\label{sec:st}

\subsection{Cross-Correlations}
\begin{figure}
\includegraphics[width=0.6\columnwidth]{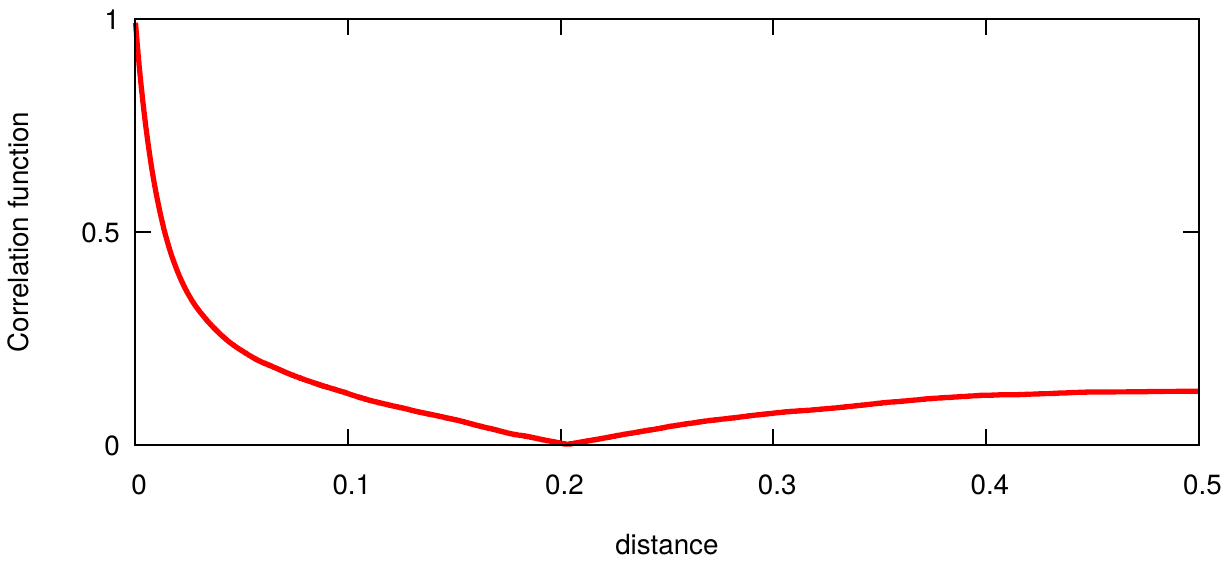}
\caption{Cross-correlations in a chimera regime with $N=32$
active and $M=8192$ passive units, calculated according to \eqref{eq:cc}.}
\label{fig:cor}
\end{figure}

To characterize the level of regularity of passive units,
we calculate the cross-correlation between the phases. 
Here, as has been shown in Ref.~\cite{Peter-Gong-Pikovsky-19},  
it is important to use a proper observable.
Indeed, because the rotations of passive phases are not free,
their distribution is not uniform -- this can be clearly seen in 
Fig.~\ref{fig:ss}, where the phases in the disordered domain are
concentrated around the value $\t\approx 2.5$. In Ref.~\cite{Peter-Gong-Pikovsky-19},
where the Kuramoto model on a STN was treated,
the transformation from the inhomogeneous phase $\t$ to a homogeneous
observable $\theta$ was performed using the local instantaneous
complex order parameter
$z=\langle e^{i\t}\rangle_{loc}$ by virtue of the M\"obius transform
\begin{equation}
\exp[i\theta]=\frac{\exp[i\t]-z}{1-z^*\exp[i\t]}.
\label{eq:mt}
\end{equation}
In the chimera setup
of this paper, we cannot properly define a local complex 
order parameter due to strong
finite-size fluctuations. Instead, we use transformation \eqref{eq:mt}
with the global order parameter of active oscillators
\[
z=\frac{1}{N}\sum_n e^{i\vp_n}\;.
\]
After the transformation \eqref{eq:mt} is performed,
the cross-correlation between passive oscillators is calculated
according to
\begin{equation}
c\left(\frac{m}{M}\right)=|\langle e^{i(\theta_k-\theta_{k+m})}\rangle |
\label{eq:cc}
\end{equation}
where the averaging is performed over all the pairs of passive phases and
over a long time interval. The latter has been chosen long enough that
every oscillator was both in regular and irregular domains. The correlation
function \eqref{eq:cc} is shown in Fig.~\ref{fig:cor}, for $N=32$ and $M=8192$.
One can see that the correlation function tends to one as $\Delta y=\frac{m}{M}$
tends to zero, what corresponds to the mentioned above continuity of the 
phase profiles.
At large $\Delta y$ the correlations are low; this is the advantage of using the
``cleansed'' observable $\theta$ instead of the original phase $\t$, for the latter
the cross-correlations do not drop below $0.4$.

\subsection{Lyapunov exponents}
\begin{figure}
\includegraphics[width=0.6\columnwidth]{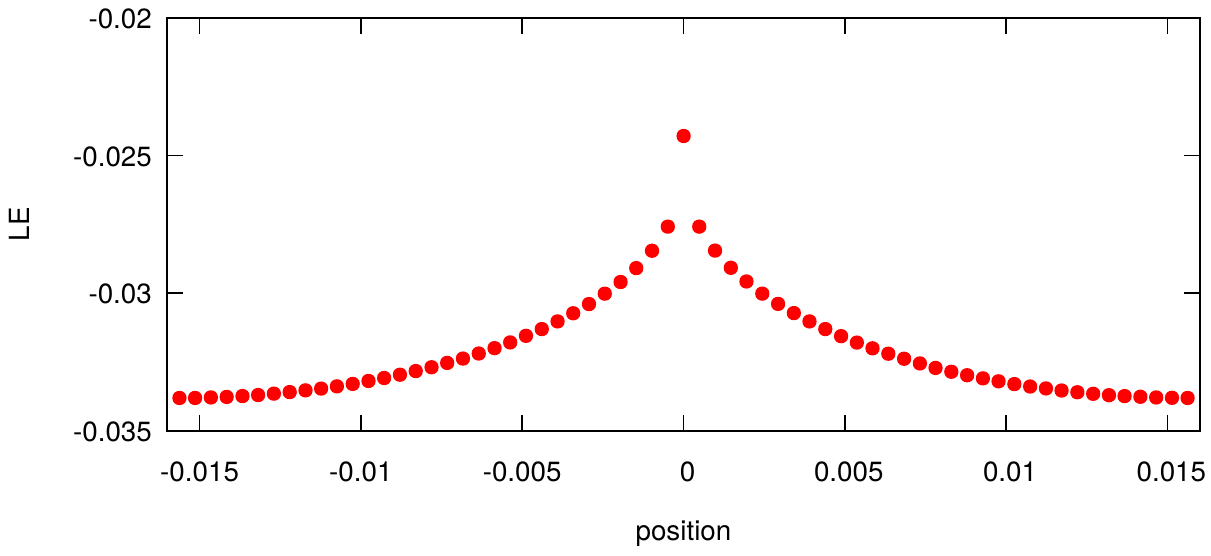}
\caption{Transversal LEs  vs position on the ring, for $N=32$. 
Due to periodicity with $1/N$, only the interval $-1/2N<y<1/2N$ around an
active unit at $y=0$ is shown.
}
\label{fig:le}
\end{figure}

In the context of STNs, there is a twofold
application of the Lyapunov exponents (LEs). Usual LEs can be defined
for a set of active particles, some of them are positive what corresponds
to turbulent dynamics depicted in Fig.~\ref{fig:vc}. For passive
oscillators, the LEs have a meaning of transversal Lyapunov exponents.
Indeed, because passive units do not act on other oscillators, the system
\eqref{eq:aosc},\eqref{eq:posc} is a skew one, and linearization of Eqs.~\eqref{eq:posc}
for passive oscillators leads to a set of independent one-dimensional equations 
for perturbations
\begin{equation}
\dot\delta\t_m=-\delta\t_m\frac{1}{N}\sum_{k=1}^N G(x_k-y_m)\cos(\vp_k-\t_m-\alpha)\;,
\label{eq:le}
\end{equation}
from which the transversal LEs (they depend on the position $y_m$), can be
expressed as
\begin{equation}
\lambda_t(y)=-\left\langle \frac{1}{N}\sum_{k=1}^N G(x_k-y_m)\cos(\vp_k-\t_m-\alpha)\right\rangle
\;.
\label{eq:leav}
\end{equation}
Calculated in this way transversal LEs are shown in Fig.~\ref{fig:le}. They are 
all negative, with the minimum at the central position between the active units.

The interpretation of the transversal LEs is as follows. If there are two passive units
at exactly the same position on the ring but with different initial conditions,
then they will eventually approach each other and synchronize. Quantity $\lambda_t$
gives the average rate of this exponential approach. In particular, if a passive unit is at the
same position as an active one, they will synchronize with the average rate $\lambda_t(0)$.
The result of this synchronization has been already discussed in Section~\ref{sec:vis}.

Negative transversal LEs explain also correlations of neighboring passive
units (Fig. \ref{fig:cor}). Indeed, neighboring sites (distance $\Delta y$) experience 
different forcing fields, therefore they cannot synchronize completely. Instead, 
one can write a model linear equation for the difference of states of passive
units
\begin{equation}
\Delta\t\approx -|\lambda_t| \Delta \t+\Delta h\;,
\label{eq:ih}
\end{equation}
where 
\[
\Delta h\approx \Delta y \frac{1}{N}\sum_{k=1}^N 
\frac{\partial G(x_k-y)}{\partial y}\sin(\vp_k-\t-\alpha)
\]
is the difference in the forcing. One can roughly estimate 
$\Delta\t\approx \Delta h/|\lambda_t|$, i.e. neighboring passive units nearly 
synchronize for small $\Delta y$. This picture is however, not exact, as the discussion in 
next section shows.

\subsection{Intermittency of satellites}
Here we focus on passive units that are extremely close to the active ones.
We call them ``satellites'', and the corresponding active unit the ``host''. 
In the Kuramoto model, such satellites are perfectly 
synchronized to the host~\cite{Peter-Gong-Pikovsky-19} (similar to the restricted many-body problem in 
gravitational dynamics, where light particles in a vicinity of a heavy body do not 
leave this vicinity). In the present chimera setup, we however observe a different 
behavior. An inspection of Fig.~\ref{fig:ss} shows that indeed in many cases
the satellites are close to the hosts (these cases a represented by blue ``lines'' passing
through red dots). However, there are at least three hosts which are 
detached from the satellites
(these are isolated red dots at $x\approx 0.075,\;0.74,\;0.97)$.

\begin{figure}
\includegraphics[width=0.6\columnwidth]{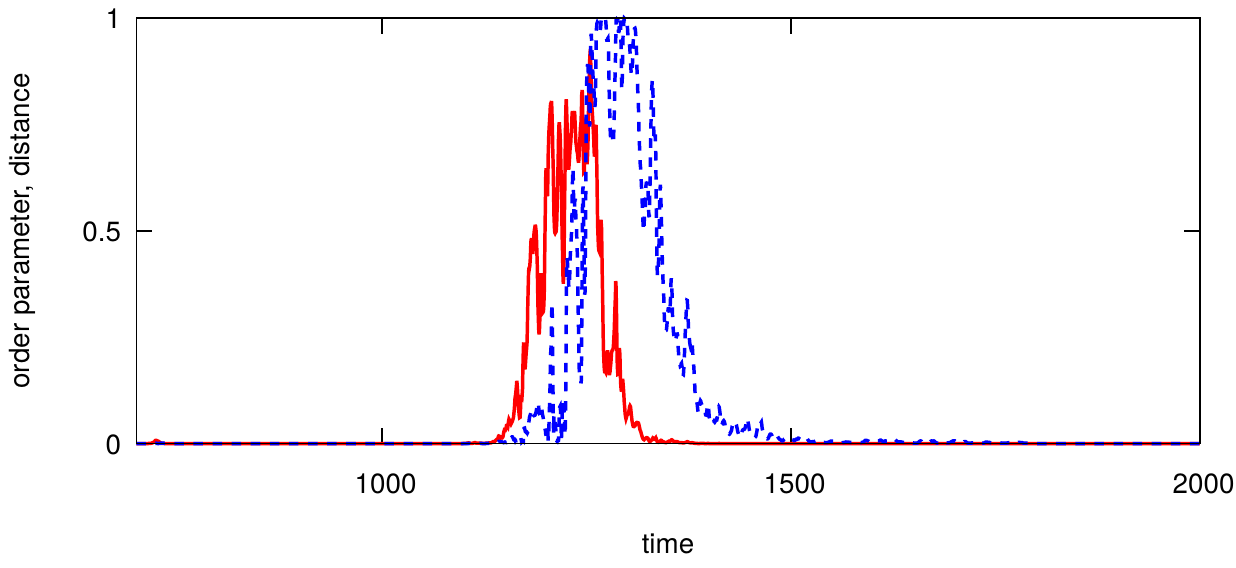}
\caption{Illustration of intermittency in the satellites' dynamics.
Red line: $1-|z_s|$; blue dashed line: distance $D$. Outside of the burst
$1-|z_s|\approx D\approx 0$. The burst has three stages. (i) first $|z_s|$ decreases 
from one, but $D$ remains small; here the satellites are spread 
around the host. (ii) Both $1-|z_s|$ and $D$ are large, satellites are spread away from host.
(iii) $|z_s|\approx 1$ but $D$ is large; satellites form a small cluster away from the host.}
\label{fig:sat}
\end{figure}

Such a behavior is not covered by the simple relation \eqref{eq:ih}. The reason is
in the fluctuations of the transversal LE, not taken into account in 
relation \eqref{eq:ih}. Such fluctuations may generally lead to so-called modulational 
intermittency~\cite{Pikovsky-Rosenblum-Kurths-01}, and this happens also here.
On average the transversal LE is negative, but the values of quantity~\eqref{eq:leav}
averaged over a finite time interval may be positive. In this case
an equation of type~\eqref{eq:ih} results in an amplification of distances, and the satellites
detach from the host. It may take a long transient time until they attach again.
This process is indeed intermittent, as Fig.~\ref{fig:sat} illustrates. In this
figure we take $L=32$ passive satellites $\t_k$ of an active host, which are spread
in the vicinity of size $- 10^{-5}\leq \Delta y\leq 10^{-5}$. To characterize
these satellites, we calculate their complex order parameter (using cleansed 
phases $\theta$)
\[
z_s=\frac{1}{L}\sum_{k=1}^L e^{i\theta_k}
\]
and depict in Fig.~\ref{fig:sat}, as functions of time, 
the absolute value $|z_s|$ and the distance from the host $\vp$ measured as 
$D=|\sin\frac{\arg (z_s)-\vp}{2}|$. Most of the time $|z_s|\approx 1$ and $D\approx 0$,
what means that all the satellites are in a small neighborhood of the host. However, there 
is a burst where the satellites spread ($|z_s|$ is as small as $0.2$) and detach
from the host. At the final stage of the burst, the satellites congregate ($|z_s|\approx 1$), 
but nevertheless remain
remote from the host ($D$ is large). 
This is a quite unusual state, which we attribute to the fact that
the transversal Lyapunov exponent is smaller in absolute value close to the host, as one can see
from Fig.~\ref{fig:le}. There is quite a long time interval 
$1300\lesssim t\lesssim 1500$, where the satellites
stay together but are detached from the host. This observation explains ``lonely active
units'' in Fig.~\ref{fig:ss}.

\clearpage
\section{Conclusion}

\label{sec:con}

In this paper we considered a special class of networks - social-type networks STNs.
From the mathematical viewpoint, they are skew systems: one active network with interconnections, which drives
another, passive network. Moreover, we assume that there no interconnections in the passive subnetwork, so that
it consists of individual driven elements. Furthermore, it is natural to
asume that the number of active elements is small, and the number of passive units is large.
This configuration mimics what really is observed in
the social networks like the facebook~\cite{Gerson_etal-17,Trifiro-Gerson-19}. We, however,
consider the effects related to STN for oscillatory systems. We have considered both
active and passive oscillators forming a symmetric ring, with long-range interactions.
Active oscillators form a chimera pattern, with a synchronous and an asynchronous domains on a ring.
Our main focus was on the dynamics of passive units. We have demonstrated that they are rather correlated,
what is explained by negative transversal Lyapunov exponents. A remarkable intermittent
dynamics is demonstrated by passive units (satellites) which are very close to an active host.
Most of the time the satellites follow the host, but there are bursts where they detach and leave the
host to move for certain time alone; after that the satellites again attach to the host.
Probably, such a behavior by followers could be observed in social networks as well.

We stress here that essential for our analysis was a rather small 
number of active oscillators. The role of this number is twofold: first, it leads 
to fluctuations of the force driving passive elements, and second, it leads to weak turbulence
of the active oscillators which restores ergodicity in the system.
Let us briefly discuss, how the effects change for large active population sizes $N$.
In this case chimera will move so slowly that the time where ergodicity 
establishes is not 
available. Thus, one should distinguish 
passive oscillators in the synchronous and the asynchronous domains. Even larger
effect
on the dynamics of passive elements is due to smallness of finite-size fluctuations.
Indeed, in the thermodynamic limit $N\to\infty$ the field acting on oscillators is stationary
in the proper rotating reference frame. Thus, passive elements will have negative LEs in the synchronous domain, 
and vanishing LEs in the asynchronous domain. The correlations, which are due to
negative LEs, disappear in this limit, and can be expected to be very weak for large
population sizes $N$.

\begin{acknowledgments}
The work was supported by the Russian Science Foundation 
(grant Nr. 17-12-01534) and by DFG 
(grant PI 220/22-1).
\end{acknowledgments}



\def\cprime{$'$}

\end{document}